# Temperature overshoot as the cause of physical changes in resistive switching devices during electro-formation

J. Meng, B. Zhao, J. M. Goodwill, J. A. Bain, and M. Skowronski

Dept. Materials Science and Eng. Carnegie Mellon University, Pittsburgh, PA 15213

Resistive switching devices based on transition metal oxides require formation of a conductive filament in order for the device to be able to switch. Such filaments have been proposed to form by the reduction of the oxide due to application of the electric field, but this report seeks to rebut that interpretation. Frequently reported physical changes during electro-formation include delamination of electrodes, crystallization of functional oxide, intermixing of electrode and oxide materials, and extensive loss of oxygen presumably to the ambient. Here we show that most of these effects are not inherent to the formation and switching processes and instead are due to an experimental artifact: the discharge of parasitic capacitances in the forming circuit. Discharge of typical BNC cables can raise the temperature of the filament to between 2,000 and 5,000 K resulting in extensive physical changes. Discharge and associated effects mentioned above can be eliminated using an on-chip load element without affecting the ability to switch.

Introduction

TaO$_x$, HfO$_x$, and related oxides are extensively used as functional layers in Resistive Random Access Memory (ReRAM) devices [1][2] with the first commercial product on the market being MN-101L 8-bit Panasonic machine control unit that used 64Kb of TaO$_x$-based ReRAM. Because of their simple structure, low power consumption, and high access speed, ReRAM has a great potential to replace traditional charge-storage-based flash memory. In addition, the arrays of such devices have been demonstrated as building blocks of neural networks [3][4][5]. These potential applications provide a motivation for improving our understanding of the processes that govern the switching characteristics of these devices.

The operation mechanism asserted to govern many ReRAM devices has been dubbed the Valence Change Mechanism (VCM) [6] which interprets switching primarily in the a chemical context. This report seeks to supplement this chemical model with an understanding of how and when thermal events can be an even more important aspect of switching. Moreover, we also examine the dynamic range of possible thermal events and how they can be controlled in switching, from events that create wholesale mayhem at the core of the switching event to ones that are more controlled and subtle.

The typical structures in this category are metal/oxide/metal sandwiches with TaO$_x$, HfO$_x$, or TiO$_x$ as the oxide and Pt, Ta, Ti, or TiN serving as electrodes. The standard (non-thermal) VCM model invokes three steps/processes needed for switching. The as-fabricated device is uniform and highly resistive and needs to be conditioned (electro-formed) to be able to switch. This is performed by application of bias which forces the oxygen ions to cross the interface with the anode leaving oxygen vacancies in the oxide [7]. The second step is the accumulation and drift of vacancies in the oxide. These defects are thought to be mobile at room temperature [8][9][10] drifting in the imposed field across the oxide layer and accumulating against cathode. After some time, the pile-up of vacancies connects the two electrodes forming a permanent conducting filament which is accompanied by a rapid drop of device resistance. The third step starts with the application of bias with opposite polarity which reverses the direction of vacancy drift and creates a gap in the filament responsible for the high resistance state.

The essential characteristics of the electro-reduction reaction in step one:

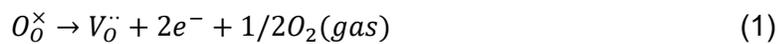

$$O_O^\times \rightarrow V_O^{\cdot\cdot} + 2e^- + 1/2 O_2(gas) \qquad (1)$$

is that it is occurs at low temperatures as the result of electric field. Before the formation of a small diameter filament, the power uniformly dissipated within the device is not high enough to significantly increase the temperature. Oxygen evolving in this process can remain dissolved in the electrode but most publications suggest it escapes forming a gas [7][11][12][13][14][15].

The strongest evidence that has been reported in support of the electro-reduction and associated oxygen evolution is the deformation of the top electrode. In some experiments, deformation takes the form of an isolated round hillock 100-1000 nm in diameter and 50 nm high [7][11][16]. The protrusion has been suggested to be caused a bubble of gaseous $O_2$ at the oxide/electrode interface. Most publications reported much more extensive deformation covering a large area up to 10 μm across with burst bubbles [12][17][18], signs of electrode melting [13], and craters reaching into the oxide layer [13][15]. These findings were corroborated by a presence of reduced functional oxide underneath the deformation [17][19][20][21][13]. An additional argument supporting oxygen evolution during electro-formation was an observation of sudden oxygen signal increase detected by Residual Gas Analyzer [14] during electro-formation.

The above interpretation, however, is at odds with earlier publications on electro-coloration and electro-degradation of oxides such as SrTiO$_3$ and BaTiO$_3$ [22][23][8]. Both phenomena were reported in bulk samples with metal electrodes at room or slightly elevated temperatures. Upon application of bias, the color and resistance of the structures slowly changed with time. The changes were interpreted as due to re-distribution of pre-existing oxygen vacancies within the sample [24][9][25]. Experiments with inert metal electrodes such as Au have not detected any loss of oxygen from the oxide [26]. This controversy needs to be addressed especially in view of an improved understanding of the electro-formation process. Among new insights into this process is the observation of threshold switching preceding non-volatile changes in devices [27][28][29], evidence of Ta ion motion during formation [30][31][32][33], and demixing of functional oxide due to lateral temperature gradients [21][31][32][34][35].

This work revisits the process of electro-reduction focusing on the runaway nature of the formation process and evaluating the temperature excursions in devices formed using different current limiting approaches. We argue that most of the gross physical changes reported to date (like bubbled electrodes, etc.) are due to the heating during the discharge of parasitic capacitances in the forming circuit and are not inherent to the electro-formation process. Chemical changes are, of-course, still relevant, to the formation process, but we assert that no interpretation of switching should be undertaken without estimating the temperature excursion. Moreover, under certain experimental conditions, which we discuss below, the temperature excursions are sufficient to vaporize sections of the device.

**Results and Discussion**

Two types of devices used in this study with lateral sizes of 2×2 μm and 500×500 nm devices were fabricated on the same chip. The bottom TiN layer used for bottom electrode and serpentine-shaped load resistor was deposited on the top of a 1 μm AlN on Si wafer (magnetron sputtering at 3 mTorr Ar pressure and 150 W power). The functional TaO$_x$ layer was reactively sputtered with the oxygen and argon flow rates of 3 sccm and 57 sccm, respectively. Pressure in the chamber was maintained at 3 mTorr. The active region of the device was defined by a via structure in SiO$_2$ layer between oxide layer and top TiN electrode. The schematic diagram of the devices is shown in Figure 1(a).

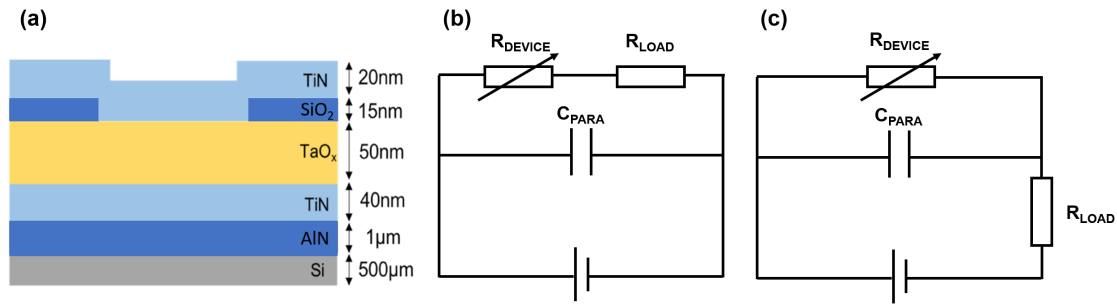

Figure 1 (a) Schematic diagram of the devices used in this study; (b) Diagram representing the circuit with the on-chip load resistor and (c) external load resistor.

Electro-formation and electrical testing have been carried out using Agilent 4155C Semiconductor Parameter Analyzer as a voltage source and ammeter. We have not used current compliance function of the instrument and relied on a load resistor of 35-40 kOhm fabricated on-chip right next to each device (Figure 1(b)) or on external one with equal resistance (Figure 1(c)). An important consideration in testing resistive switches is presence of parasitic capacitances in the circuit. Even though the testing is typically done by quasi-static current or voltage sweeps, the rapid change of device resistance upon forming or switching can lead to capacitive discharge accompanied by current and temperature excursions affecting device characteristics [36][37][38][39]. ReRAM structures are particularly vulnerable to current spikes as the energy is dissipated in a very small volume of the conducting filament. In the circuits in Figure 1, the capacitance $C_{PARA}$ represents the capacitances of leads connecting the device to the sourcemeter, micromanipulators, contacts pads, etc., the biggest of them, in typically used experimental systems, being capacitance of the BNC cables (80 pF/M). In experiments we have used 1m long cable giving $C_{PARA}$ of 80 pF. In simulations, we have used three values of $C_{PARA}$ of 80, 20, and 0 pF. The external load resistor, when used, was located on the sourcemeter side of the cable (Figure 1(c)). This circuit mimics use of the sourcemeter as a current source employed in many experiments. It is easy to notice that in the circuit shown in Figure 1(b), the capacitor is directly connected to the terminals of the voltage source. Since the voltage was either constant or changing slowly in all experiments and simulations, the circuit in Figure 1(b) is equivalent to one with no capacitor.

Electro-formation experiments were performed by quasi-static source voltage sweeps from 0 V to 13 V. Figure 2 shows the *I-V* characteristics of two representative 500 nm devices tested with on-chip (a) and external (b) 35.6 kOhm load resistors. Please note that all *I-V*'s are plotted as a function of voltage drop across the device rather than the source voltage. Electro-forming traces (blue curves) start with current increasing superlinearly with voltage. At about 10 V, both *I-V*'s make a rapid transition to the *I-V* of a formed device. The transition is marked by a dashed line connecting the last point in the unformed *I-V* and the first measured point in the electroformed device. There are no datapoints during the transition as the data are collected by the sourcemeter at time intervals much larger than the time needed to form the device. In the circuit without parasitic capacitance, the transition follows the load line. In the presence of capacitance, the transient *I-V* lies above the load line [40]. As is typical of all our fabrication and testing procedures, devices formed to the high resistance state with highly non-linear characteristics [41][28][32][42]. Both as-formed *I-V*'s exhibited a negative differential resistance region at currents above 30 µA. The device formed with on-chip resistor had higher resistances at low voltages.

Devices were switched to the low resistance state by the application of negative bias of 1-1.5 V with roughly linear *I-V* trace (red traces in Figure 2(a) and (d)). This was followed by a typical *I-V* switching characteristics (not shown). The differences between *I-V*'s of two devices are minor.

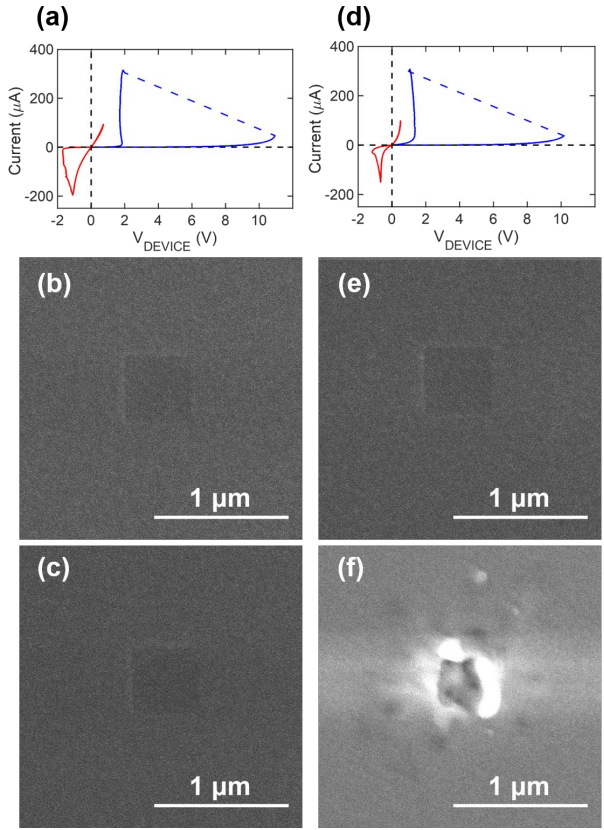

Figure 2(a) Electro-forming and switching quasi-static *I-V* characteristics of a 500 nm device with the on-chip 35.6 kOhm load resistor. (b) and (c) SEM images of the top electrode morphology before and after testing of device in (a). (d) I-V characteristics and (e), (f) SEM images of device tested in circuit with external load of 35.6 kOhm.

Devices used in electrical testing have been imaged by Secondary Electron Microscopy (SEM) before and after testing. All images show the top TiN electrode covering the entire field of view. The active part of device corresponds to the square 500x500 nm via etched in the $SiO_2$ layer between $TaO_x$ and top electrode. Panels (b) and (c) in Figure 2 depict device formed with the on-chip load resistor. There are no apparent changes to the electrode morphology. This is in agreement with SEM and TEM analyses of several hundreds of devices of different sizes, $TaO_x$ layer thickness and composition, substrate material, and testing conditions performed in this group. Devices formed and switched with an on-chip load resistor did not show changes of electrode morphology. Figures 2(e) and (f) show corresponding before and after images of nominally identical device tested with the external load. It is apparent that the morphology of the device has undergone major changes during forming. Most the TiN from the active area has peeled off with accumulation of material at the edges of the crater. Very likely part of the device area can no longer contribute to conductance due to lack of electrical contact. Despite this, the device *I-V* characteristics do not seem to be affected. Apparently, the as the top electrode was separating, the current flowing though the device formed the part that still made to connection with the electrode.

An example of a localized "bubble" created during forming is shown in Fig. 3. Panels (a)-(d) correspond to a device tested in a circuit with on-chip load of 38.8 kOhm while panels (e)-(h) represent device tested with external load ($R_{LOAD}$=38.4 kOhm)). The forming and switching characteristics for both devices are quite similar with the "knee" of the S-type Negative Differential Resistance characteristics of as-fabricated device at 9.5 V [29] and snap back to a formed high resistance state [42] followed by standard switching *I-V*. Similarly as in Figure 2(b) and (c), there are no visible morphology changes of the device. Corresponding bright field transmission electron microscopy image (Figure 3(d)) shows AlN, TiN, and $TaO_x$ layers with flat interfaces with roughness of approximately 5 nm due mostly to rough AlN surface. The grey area above the top TiN electrode corresponds to protective Pt layer deposited in the Focused Ion Beam system after testing of the device.

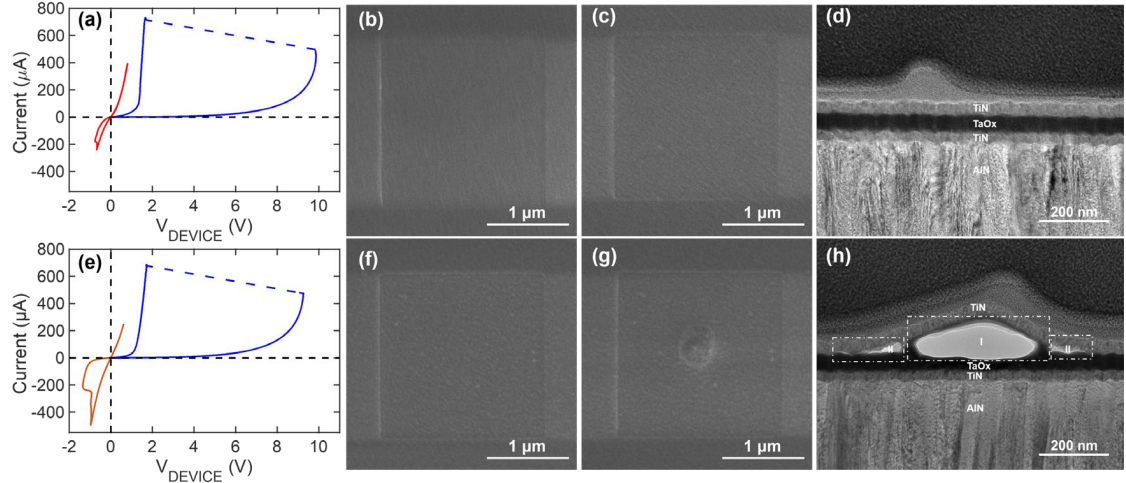

Figure 3(a) Forming and switching characteristics of a 2×2 μm device using a circuit in Fig. 1(b). (b) and (c) SEM images of device before and after testing. (d) Bright field TEM cross-section of the deformed area in (c). Panels (e)-(h) are corresponding I-V and images obtained on device tested using a circuit with on-chip load resistor.

The SEM image in Figure 3(g) of the device formed with external load shows a new feature created during electro-formation. It is a circular hillock with a diameter of approximately 450 nm located in the center of the via. Our devices typically form in the center as evidenced by the TEM images in our recent publication [32]. Figure 3(h) is a bright field transmission electron microscopy image of the device cross section in the area of the hillock. Compared with the reference TEM image in Figure 3(c), the image shows three voids. The central one corresponding to light grey oval in the middle of the image, has diameter of 400 nm and height 120 nm. The void is located in the upper part of $TaO_x$ layer visible as the darkest layer in the structure. It is interesting to note that the central void is not at the oxide and TiN interface with the thin dark Ta containing layer visible at the top of the void. In contrast, the two triangular voids on the sides of the figure (marked by "II") are located at the $TaO_x$/TiN interface. This suggest that the mechanisms of formation for the central and side voids are different. Below we will argue that the central void forms at the location of the highest temperature in the entire device and it forms first. The side voids appear to be due to delamination of the top electrode caused by thermo-mechanical stresses.

Similarly as in case of 500 nm devices, large physical changes of the device structure have not visibly affected the *I-V* characteristics. This should not be surprising as the area of the voids effectively eliminates only about 5% of the device area. Moreover, current is mostly flowing through the small filament located somewhere in the structure and, based on statistics, unlikely to be found in the TEM sample.

The conclusion of the experimental results described above is that the formation of bubbles, voids, or craters in the resistive switching devices is not intrinsic to electro-formation or resistive switching processes. Instead, in most instances it is associated with lack of current control in circuits not using integrated load elements such as resistor or transistor. Discharge of parasitic capacitances always present in all circuits is apparently causing large excursions of current and temperature responsible for formation of voids and/or electrode delamination.

The critical parameter in formation of voids in switching structures is the maximum temperature reached by the device during such discharge. Below we argue that the temperature excursions can be very large leading to major physical changes in the device structure. One can start with a back of the envelope estimate of the temperature increase. Let us assume that the filament in the switching device is 100x100x100 nm (a very conservative estimate) and device forms at bias of 5V. If the energy stored in the 80 pF capacitor is dissipated within the filament adiabatically, the filament would reach temperature of approximately $7 \times 10^5$ K. In other words, the energy is high enough to vaporize the entire filament indicating that the effects of discharge need to be carefully considered.

A more precise temperature estimate was produced using a finite element model which solved coupled charge and heat flow equations with an additional restriction imposed by the circuit. Only the circuit with the external load needed to be considered as the on-chip load configuration in Figure 1(b) is equivalent to circuit in Figure 1(c) with $C_{PARA}$=0 pF. Detailed description of device structure, governing equations, material parameters, and boundary conditions can be found in the Supplementary Information. Voltages and current and temperature distributions were calculated as a function of time in response to a constant source voltage pulse $V_{SOURCE}$=5V. This value was selected as typical of the forming voltage and energy stored in capacitance of the cables in many experiments reported in the literature. Initially, the TaO$_x$ layer was assumed to be uniform with the conductivity given by Frenkel-Poole formula with pre-exponential factor of 1400 and activation energy of 0.35 eV [29]. The device was allowed to reach the steady state current and temperature distribution by letting the circuit evolve for 100 µs. Until this point, the device acted as a simple resistor. At 100 µs, a new element was introduced in the model. In order to mimic the electro-formation process, we have assumed that a cylindrical metallic filament with the diameter of 16 nm appeared in the structure. This size was selected to reflect the most reliable estimates of the filament diameter [32][43][44]. Its initial temperature was that of the steady state temperature of the center of device. The filament conductivity was assumed to be independent of temperature and has linearly changed from 0.6 S/m to 3x10$^4$ S/m over time period $\tau_{EF}$. The upper limit of filament conductivity was selected to reproduce experimental value of resistance of the 2 µm device at the highest point of I-V right after electro-formation. The characteristic time of electro-formation ($\tau_{EF}$) is a strong function of applied bias [45][42]. We have treated it as a parameter and calculated the circuit response for $\tau_{EF}$ of 20 ns, 200 ns, and 2 µs in order to span part of the range of reported values. Similarly, we have calculated current and temperature transients for three different values of parasitic capacitance of 0, 20, and 80 pF.

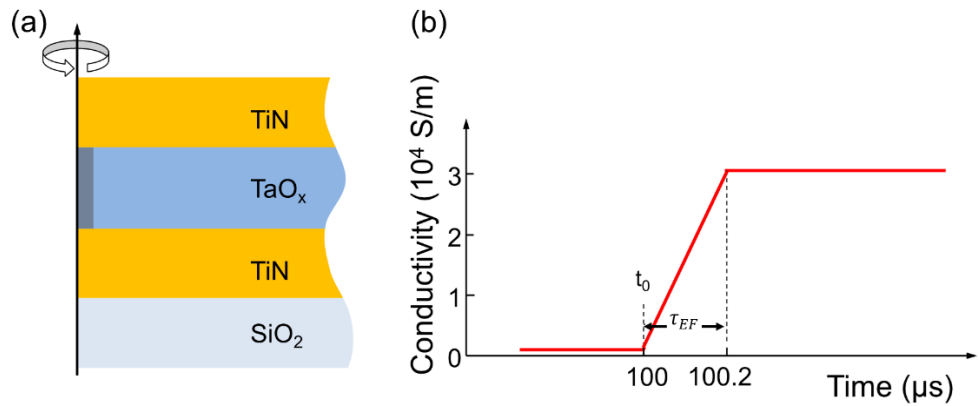

Figure 4. (a) Schematic diagram of the filament in the device structure. (b) Conductivity change of the filament versus time selected to mimic the electro-formation process.

Simulation device current as a function of time for the $\tau_{EF}$=200 ns is shown in Fig. 5(a). Black trace for $C_{PARA}$=0 pF shows a monotonic current increase as the filament forms and is constant afterwards. Corresponding temperature transient in Figure 5(b) starts at temperature of a little above 300K has a maximum of 3750 K at 200 ns after $t_o$ and stabilizes at 450 K at long times. The appearance of maximum is due to $V_{SOURCE}$ division between the device and the load. Before electro-formation, the device resistance is much higher than the load while after the electro-formation it is considerably lower. Initial $V_{DEVICE}$ is close to applied bias and drops to 0.886 V after the transition. The power disspated in the device has a maximum when the device resistance is equal to that of the load which happens when electro-formation is 80% complete. The same current and temperature transients as the black trace are obtained if using on-chip load for any value of $C_{PARA}$. Red and blue traces in Figure 5 (a) and (b) correspond to capacitance of 20 pF and 80 pF. Current starts increasing with the same slope as for 0 pF but increases for significantly longer time due to extra current supplied by the capacitor. Significant fraction of charge stored in 20 pF capacitor discharges during electro-formation causing a near saturation of current. For 80 pF, current increases almost linearly throughout the transition. Once the device resistance stabilizes, the current decreases exponentially with the time constant [46]:

$$\tau = \frac{R_{DEVICE} \times R_{LOAD}}{R_{DEVICE} + R_{LOAD}} \times C_{PARA} \qquad (2)$$

The temperature transients closely follow the dissipated power which is mostly determined by current. The drop of $V_{DEVICE}$ is reflected in a maximum appearing in red trace and larger deviation from linearity for the blue one. Highest temperatures reached in the middle of the filament are 3750 K, 2160 K, and 550 K for capacitances of 80, 20, and 0 pF.

Current and temperature transients for a faster electroforming process ($\tau_{EF}$=20 ns) look quite similar to the ones for $\tau_{EF}$=200 ns. Black trace is not affected at all other than the change of the time scale. The current and temperature rises for $C_{PARA}$ of 20 and 80 pF are almost identical as there is little discharge in a short time of electro-formation. The decay rate for 20 pF being 4 times as fast as the one for 80 pF. The maximum temperatures are a little over 4,000 K. For a slower formation process ($\tau_{EF}$=2 µs), charge stored in the capacitor is mostly removed with the current transients for 20 and 80 pF showing a maximum during the transition. This results in significantly lower maximum current and temperature with $T_{MAX}$ of 980 K and 1550 K for 20 and 80 pF, respectively.

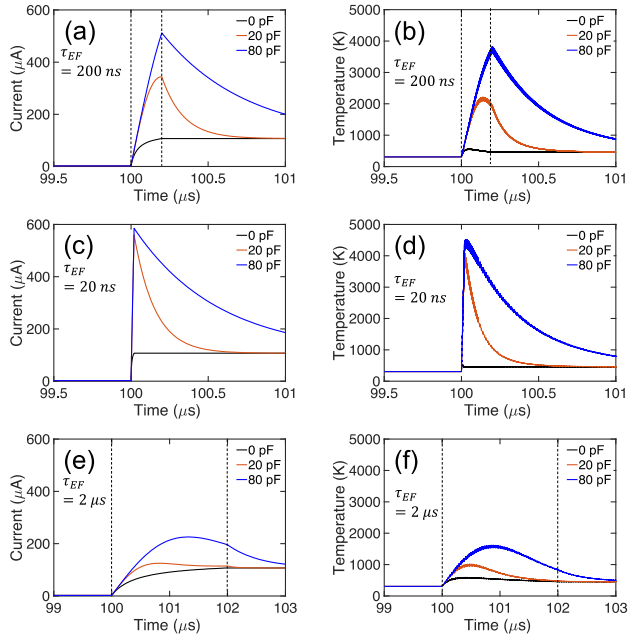

Figure 5 (a), (c), and (e) show current transients for electro-formation times of 200 ns, 20 ns, and 2 µs, respectively. Panels (b), (d), and (f) illustrate change of temperature in the middle of the filament for the same electro-formation times. All traces were calculated for an external load resistor shown in Fig. 1(c). Solutions for the case of an on-chip resistor are identical with the black curves for any value of $C_{PARA}$.

Discussion

The simulations above indicate that the discharge of the parasitic capacitance can increase the temperature of the conducting filament by several thousand Kelvin. The highest temperatures are expected within functional oxide layer close to the interface with the anode in devices formed in positive polarity [32]. The SEM and TEM images are consistent with this result: the deformation of top electrode appeared only in devices that experienced the discharge with the major void located in the upper portion of $TaO_x$ layer. While it clearly is related to the high temperatures, the responsible mechanism can only be speculated about. Boiling point of $TiO_2$ was estimated at

2,500-3,000 °C with that of $ZrO_2$ and $HfO_2$ at 4,000-5000 K [47]. The boiling point of $TaO_2$ used in this work should be toward the lower end of this range. It is conceivable that the temperature within the oxide exceeded the boiling point producing a bubble of $O_2$, TaO, $TaO_2$, and other radicals [48]. Alternatively, the voids could be forming due to local melting and thermal expansion of materials. The weakest bonds in resistive devices are between the functional oxides and noble metals such as Pt frequently used as electrodes. In these structures, the bubbles would be expected to form by electrode delamination due to thermal expansion of heated part of the device clamped to silicon substrate at 300 K. The two side-voids in Figure 3(h) appear to be due to such process.

The conditions used in the simulations have been selected to correspond to commonly used electro-forming and switching conditions. Many results concerning the structure of the filament and active mechanisms of formation were obtained using current compliance feature of the voltage source [17][21][49][50][51][52][53][54][55][34][16]. Most of the remaining reports have not mentioned any current limiting approaches. Neither current compliance not use of the current source can prevent the electrostatic discharge of parasitic capacitances. The often reported reduction of the functional oxide does indeed occur but it could be entirely due to the large temperature excursions during discharge.

It needs to be pointed out that the high temperatures are not necessary for the electro-formation or switching. The capacitive discharge and associated effects are artifacts of improperly conducted experiments. Unfortunately, these artifacts are frequently interpreted as occurring in all switching events and observed changes as typical of switching. The discharge effects had a particularly high impact on the interpretation of the filament structure. Microscopy techniques such as TEM and x-ray spectro-microscopy require samples in a form of a thin lamella which are typically prepared from the locations of the electrode deformation [17][55][14][16][13][20]. The reasoning is that "the most damaged area is likely to give more detailed information on the breakdown mechanism of devices". If devices were not properly protected, the most damaged area will also produce most artifacts and most dubious interpretations. The approach advocated here, i.e. on-chip current limiter is not a new solution and it is not a perfect one either. It does not eliminate the discharge of the device self-capacitance and even though this process dissipates orders of magnitude less energy, it still affects the device characteristics [56].

While the above disputes the use of electrode morphology changes as an argument for electro-reduction of functional oxides in switching devices at low temperatures, it does not eliminate the possibility of electro-reduction altogether. However, one can easily notice that if the oxygen vacancies or other defects that control the conductivity in the switching device can move at room temperature, then the filament consisting of such defects should dissolve in the time comparable to the time needed for electro-formation. Large concentration gradients would drive the Fick's diffusion and homogenize the material. This would defeat one of the primary potential applications of switching devices. Memories requiring 10 year retention cannot be based on defects with significant diffusion coefficient at room temperature unless there is a locking mechanism preventing defects from leaving the filament.

Conclusion

In this report we have re-evaluated the effects of forming procedures widely used for conditioning of resistive switching devices. We have argued that many currently accepted interpretations have been arrived at by using questionable experimental results stemming from the poor control of the current excursions during forming. Specifically, the electric-field-induced oxygen loss thought to initiate the filament formation appears to occur at high temperatures, possibly being the consequence of the filament formation rather than its cause. Most of the reported physical changes in the device structure including extensive crystallization of the oxide, reduction of the oxide due to oxygen loss to ambient, and delamination of electrodes have been obtained without proper current control and should be discounted as part of the electro-formation and switching process. In short, we need to re-evaluate most of the interpretations of the structure of the filament and the mechanism of switching offered to date.